# A Strategic Cyber Crime and Security Awareness Information System using a Dedicated Portal

**Moses Adah Agana[1], Bassey I.Ele[2]**

[1]**mosesagana@unical.edu.ng, +2347036673324,** [2]**mydays2020@yahoo.com, +2348064451381**

[1,2]**Department of Computer Science, University of Calabar**

**Abstract:** A real time portal (www.ganamoscybersecure.org) to enlighten people on how to protect their data in the web, the strategies adopted by cyber criminals to succeed in exploiting their victims as well as the mistakes made by people and organizations that make them prone to the menace of cyber crime has been designed in this paper. Critical observations were made over time on the degree of awareness and attitudes of people and organizations in some parts of South Africa and Nigeria towards information security. Interviews were conducted using structured questionnaire to elicit information. The outcome showed that most individuals and organizations lack adequate awareness on strategies adopted by cyber criminals, and pay little attention to securing their online data. The application of the services provided by the portal designed would be a sine qua non for contending with the menace of cyber crime.

# 1. Introduction

Information security is really about business continuity while taking precautionary measures against natural disasters, criminal attack, and errors committed by users or caused by system failure. The Internet and related technologies are prone to a variety of crimes orchestrated by a variety of persons or groups [1]. The increasing dependence of businesses on computer systems and the Internet has made many more organizations vulnerable to the impact of cyber crime. Indeed, more companies are worried about the risk of cyber crime than they are about product liability, fraud and theft.

Automated tools became inevitable for protecting information and data because of the ubiquitous and virtual use of the computer for processing, storing and communicating data and information. The non-physical nature of automated tools used in information security has created opportunities for criminals to anonymously misuse information especially on the Internet [2]. Any crime committed using a computer and a network of telecommunication devices is termed cyber crime. The computer may be the target or it may be used in committing the crime [3].

Cyber crimes can threaten a nation's security and financial health. Computer system vulnerabilities persist worldwide, and initiators of the random cyber attacks that plague computers on the Internet remain largely unknown. Conventional crimes such as armed robbery, rape, stealing, cultism, etc. are easy to detect and the culprits can be prosecuted by law enforcement agents because of their physical nature [4]. This is not the case with cyber crime due to the fact that the cyberspace is quite wide and is a virtual reality. The cyberspace has become an environment where the most lucrative and safest crime thrives. This is partly due to lack of adequate awareness by Internet users, inadequate security restrictions to Internet access, lack of proper cyber user identification/detection techniques and loose cyber regulations on prosecution of the culprits.

Most organizations and individuals are oblivious of the tricks adopted by cyber criminals to exploit them, and equally seem to pay less attention to strategies of securing their online data from cyber attacks. A major step to overcome this growing trend of insecurity in the cyberspace is to develop a strategy to educate individuals and organizations on the risks of information insecurity and how to avoid falling prey to the cyber criminals.

*1.1 Scenarios of Insecurity in Data Transmission*

Some common scenarios of insecurity in data transmission across communications networks abound at all times. They include the following [2] [5]:
   i. User A transmits a file containing sensitive information (like a bank deposit) to user B that needs to be kept secret, but user C who has no authorization to read the file monitors and captures a copy of the file during transmission. He could divulge the information or use it for something else.
   ii. A network manager X transmits a message to a computer Y under its management so that the computer (Y) can update a file authorizing the inclusion of the identities of a given number of new users who are to be granted access to the computer. A certain user Z intercepts the message and alters its contents to add or delete entries before forwarding the message to computer Y. Computer Y will ignorantly accept the message as originating directly from X and act on it without knowing that it was modified by Z on transmission.
   iii. A variant of (ii) above could be a situation whereby user C will create a message in the name of manager X and send it to Z; and Z will act on it, unknowing that it was C that impersonated manager X.

iv. In another instance, an employee is fired without warning for misdemeanor and the personnel manager sends a message to a network server to disable the employee's account and post a notice to the employee's file as a confirmation of action after successfully disabling the account. The employee is however able to intercept the message, delay it sufficiently enough to enable him access the server and retrieve some vital information which he may use to mar the organization after he has been sacked.
v. The recipient of a message could deny receiving the message or the sender of a message could deny sending the message (repudiation).

An inexhaustible list of such scenarios abounds as cases of information security and cyber crime today. The above scenarios involve the use of codes to attack information on networks, and are more difficult to detect and manage than physical attacks that involve the use of certain weapons such as knife, fire, bombs, and hammer to destroy computer installations and networks.

*1.3 Strategies adopted by Cyber Criminals*

Any mission, whether authorized or unauthorized cannot succeed without careful planning and execution. Cyber criminals plan over a long time in other to succeed; and they use hacking techniques which the Internet itself provides. Five main steps are adopted by cyber criminals to hack into and access their victim's computer, take control and achieve their objectives [4] [6] [7] [8]. These steps can be summarized as follows:

i. The hackers first spy their target, and possibly trick them to reveal some confidential information about their organization. This is termed *Reconnaissance and Pre-operative Surveillance*.

ii. The second step is that the hacker further spies the victim's computer or network for possible loose ends to gain access without security restrictions. This can be achieved by "war dialling" (continuously dialing the victim's line until he responds, then he gets him) or "war driving" (driving around the victim's network until an opening is located for entering to commit crime).

iii. The third step is gaining access, which the attacker achieves by using a password he has guessed or stolen to create a phony account, or he exploits a vulnerable point to let him install some malware to use for further attacks in the network.

iv. The fourth step is that the attacker maintains access after successfully gaining unauthorized access; by installing some malware to enable him completely take over the network as the administrator. He can thus block other hackers from hacking into the network by installing software patches that can close the previous security vulnerabilities.

v. The fifth step is covering of tracks. In order to maintain control, the hackers after taking over a network cover all tracts that may provide access to other hackers and try to secure all vulnerabilities with the existing system. The Internet provides new "anti-forensics tools" that hackers can use to conceal their actions from cyber crime investigators.

*1.4 Reasons why People fall Prey to Cyber Attacks*

A good number of reasons account for why people and organizations fall prey to cyber attacks, prominent among these is the increasing dependence on computer networks. Operating an unsecured or inadequately secured computer network can expose a business to a number of risks [9]. When networks are poorly secured, possibilities exist for criminals to exploit. The open source nature of the Internet makes it easy for anyone to easily access some confidential records of an organization. A business opponent who has a clue to such useful records of the other can use the information to threaten the success of the rival.

Vulnerable unidentified weakness in the two-step authentication/verification process as used by most banks whereby customers receive changing codes by mobile phone to use with their passwords can also expose users to the risk of cyber crime. The attackers exploit this vulnerability to hijack the legitimate user's session and break into his bank account to swiftly transfer funds to their own accounts unnoticed. An instance of this scenario was identified by Kaspersky, the Russian cyber security company where more than 100 banks - mainly in Eastern Europe were raided by cyber criminals as a result of the vulnerability [10].

In a related development, it has been reported that hackers are exploiting the Signaling System 7 (SS7) to hijack sessions of bank customers and sweep their funds unnoticed [11]. SS7 is an international telecommunications standard (the backbone of worldwide mobile communication) that specifies how cell phone networks connect with each other and allows cell phone users to roam on networks anywhere else in the world if they are out of their country of operations of local telecommunication services. With SS7, users can make and receive calls, as well as text messages across networks even if out of their country. Once the hackers have gained access to the SS7 network, they can impersonate a phone's location, read or redirect messages, and even listen to calls and this is a vulnerable security situation [11]. Even the network-initiated Unstructured Supplementary Service Data (USSD) that is used for authenticating transactions in the scheme can still be illegally hijacked or redirected just like calls because it also depends on the handset's SIM card.

The inability of individuals and organizations to understand social engineering can also make them susceptible to cyber attacks. Social engineering is concerned with the tricks adopted by hackers to create a near replica of a system to convince innocent users of the Internet to divulge their confidential information like passwords, personal identification numbers (PINs), bank account numbers, etc. to them. As an instance, most criminals put up phone calls to an innocent person, may be a company's representative claiming to be an agent or customer representative of a certain company that offers Internet services, convincing him to release some sensitive information like his password without checking the genuineness of the call. The dubious caller uses the information to exploit the person's or organization's web resources.

Russian cyber criminals have been reported to have utilized social engineering to exploit vulnerability in Telegram's desktop instant messaging client for Windows to deliver malware to users from March 2017 using the advantage of the fact that the Telegram app for Windows accepts and uses a specific character [12]. The criminals were said to have prepared the malware (a JavaScript file) and gave it a name and icon quite similar to what the users know without raising suspicion.

Failure to update an operating system to a current version with enhanced security can also expose a system to the risk of cyber attack. Chinese cyber criminals have been reported to have exploited the vulnerability of Windows XP (due to its low security provisions) using the software called **Rufus** to access ATM cash dispensers without being noticed (in some regions of the country - Odisha, West Bengal, Bihar and Gujarat [13].

*1.5 First-hand Precautions against Cyber Attacks*

Access control through the use of passwords, personal identification numbers, etc. is an efficient way of contending with hacking [14]. Cryptography (encryption) of data transmitted via networks also secures them from hackers [15]. Other strategies that can be adopted by individuals and organizations to prevent their information from falling prey to hackers are business continuity and disaster recovery planning [16], as well as information security governance and risk management [17].

As a first and paramount step, individuals and organizations need to fortify their data from cyber attacks if they:

i.   Use strong passwords, do not reveal passwords, do not use the same passwords for different accounts and enable 2-factor authentication for their webmail, if possible.
   ii.  Always log out of accounts after session, and apply patches or software updates.
   iii. Understand where their data and information reside and how to categorize them.
   iv.  Encrypt sensitive data sent across the cloud [18].
   v.   Understand and apply social engineering.

To completely avoid the kind of eavesdropping prevalent in SS7, opening a completely isolated, end-to-end encrypted communications channel between the mobile phone and the servers that process payments or store sensitive data, and to properly authenticate the users of such channel is recommended [11].

For business continuity and disaster recovery planning as well as risk management, strategies that individuals and organizations need to adopt to forestall letting their data and information prone to cyber attacks include:
   i.   Engaging information technology professionals for periodic system upgrades.
   ii.  Training of employees on information security, reporting suspicious security actions and holding them accountable for information security breaches.
   iii. The use of strong firewalls.
   iv.  When hiring, criminal background checks should be conducted to avoid employing people that will become internal threats to the information resources.

## 2. Objectives

The main aim of the study is to develop a cyber crime and security awareness information system to contend with the menace of cyber crime.

The specific objectives of the study are to:
1. Review the strategies adopted by criminals in exploiting their victims
2. Determine the level of awareness of individuals and organizations on cyber crime strategies and their attitudes towards information security.
3. Use the information gathered to develop a real time cyber security wareness portal to aid individuals and organizations in avoiding falling prey to cyber criminals.

## 3. Methodology

Two methods were adopted for the system design:

*Method I:* A blind survey was conducted to elicit facts about the level of awareness and attitudes of people and organizations towards cyber security. Critical observations were also made by visiting some organizations that provide financial and related services mediated over the Internet to the public. The identities of those interviewed as well as the organizations contacted were not sought for in order to avoid infringing on their security rights to enable freedom of expression without being held liable for divulging some critical information. The research was conducted in Harrismith and Phuthaditjhaba in Free State, South Africa, and in Calabar, Cross River State, Nigeria.

Two questionnaires requiring YES/NO responses were administered in each of the areas; one was to test the degree of awareness of individuals and organizations on cyber security while the other was to test their attitudes towards the security of their data (whether they show laxity or seriousness). Each response option with the highest number of respondents was considered as the opinion to uphold. The respondents were randomly selected. 7 organizations (4 in South Africa and 3 in Nigeria) as well as 50 individuals (30 in South Africa and 20 in Nigeria) constituted the targeted population.

*Method II:* The iterative method of system analysis and design was adopted in the development of the cyber security portal. Requirements were reviewed, designed, tested and implemented and as new requirements evolved, the process of analysis and design is reviewed, followed by a new system design, testing and implementation. The design was achieved using Personal Homepage Preprocessor (PHP) as the scripting language.

## 4. Results/Discussion

The results of the study are presented in two folds: first the outcome of the interview conducted and second, the cyber security awareness portal designed.

*4.1 Interview Results*

From the questionnaire to determine the level of awareness of individuals and organizations on cyber security, representatives of 5 out of the 7 establishments contacted representing 71.4% were unaware of how insecure their data and information are in the cyber space due to the way they handle transactions and the activities of cyber criminals while only 2 out of 7 representing 28.6% were aware.

Similarly, 33 out of the 50 individuals contacted representing 66.0% were unaware of how insecure their data and information are in the cyber space due to the way they handle transactions and the activities of cyber criminals while only 17 out of 50 representing 34.0% were aware. This implies that there is a great need to create awareness on information security strategies for the organizations and individuals. The outcome of the responses to this questionnaire is summarized in table 1.

*Table 1:* Responses showing level of awareness individuals and organizations have about cyber security

| Category | Percentage aware | Percentage unaware |
|---|---|---|
| Organizations | 28.6 | 71.4 |
| Individuals | 34.0 | 66.0 |

Figure 1 shows the pie chart representing the levels of cyber security awareness by organizations that were examined and figure 2 shows the pie chart representing the levels of cyber security awareness by the individuals that were examined.

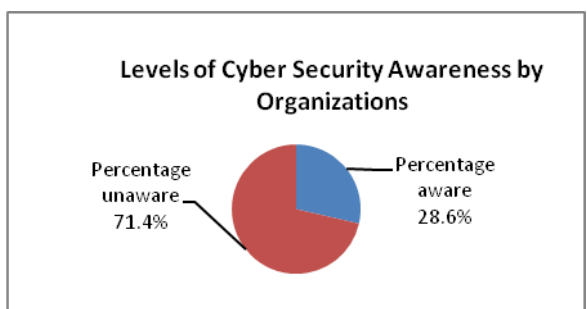
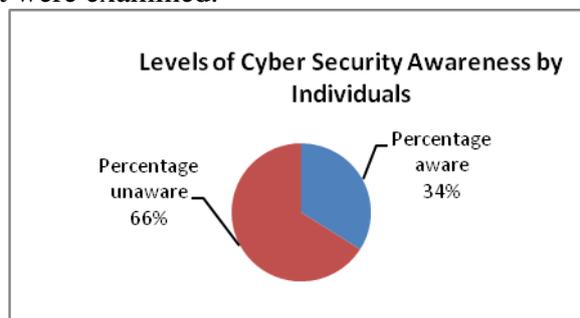

Figure 1: Levels of Cyber Security Awareness by Organizations

Figure 2: Levels of Cyber Security Awareness by Individuals

From the questionnaire to determine the attitudes of individuals and organizations towards security of their data, representatives of 3 out of the 7 establishments contacted representing 42.9% proved that they show serious concern for the security of their data while only 4 out of 7 representing 57.1% proved that they show little commitment to the security of their data in the cyber space.

Similarly, only 15 out of the 50 individuals contacted representing 30.0% proved that they show serious concern for the security of their data while only 35 out of 50 representing 70.0% proved that they show little commitment to the security of their data in the cyber space. This implies that there is a great need to create awareness on the need to commit reasonable resources to information security strategies for organizations and individuals. The outcome of the responses to this questionnaire is summarized in table 2.

*Table 2:* Responses showing attitudes of individuals and organizations towards the security of their data

| Category | Percentage showing serious commitment | Percentage showing |
|---|---|---|

|              |      | little commitment |
|--------------|------|-------------------|
| Organizations| 42.9 | 57.1              |
| Individuals  | 30.0 | 70.0              |

Figure 3 shows the pie chart representing the levels of commitment of organizations towards the security of their data, while figure 4 shows the pie chart representing the levels of commitment to data security by individuals.

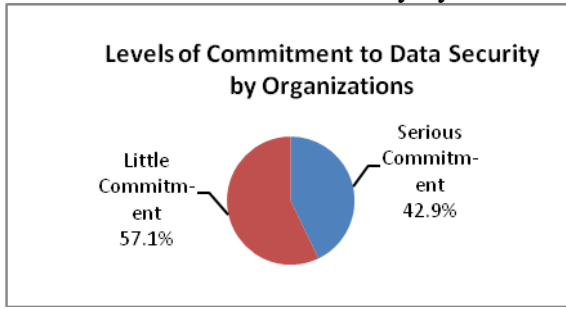 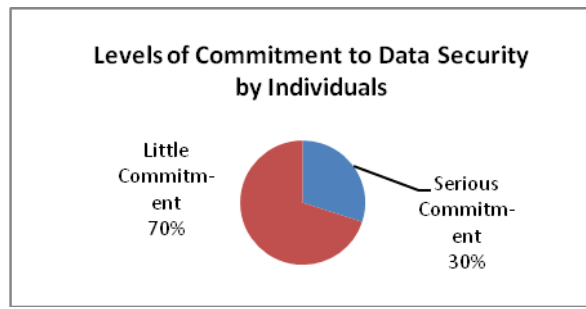

Figure 3: Levels of Commitment to Data Security by organizations

Figure 4: Levels of Commitment to Data Security by organizations

*4.2 Cyber Security Awareness Portal*

The features of the cyber security awareness portal developed are briefly presented in this section. The portal is interactive, offering real time cyber security tips to users. It is launched by typing the URL www.ganamoscybersecure.org in the address bar using any web browser.

Figure 5 shows the home page of the site from where the user can navigate to various pages, while figure 6 shows the contacts page for cyber security services.

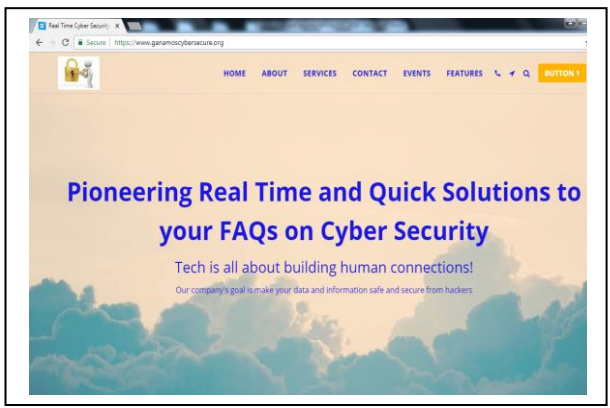 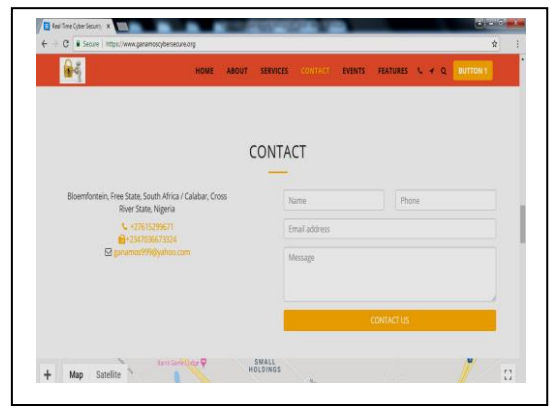

Figure 5: Home Page of the Cyber Security Awareness Portal

Figure 6: The Contacts Page

In figure 6, clients can contact information technology professionals through the site. This they can do either through direct phone calls or email messages. Immediate feedbacks are provided once messages are received from clients.

The most fascinating of all the pages is the features page that offers link to various tips on how to protect data and avoid falling prey to hackers. This is illustrated in figure 7.

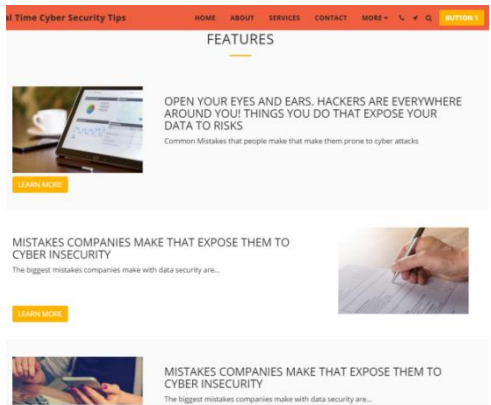
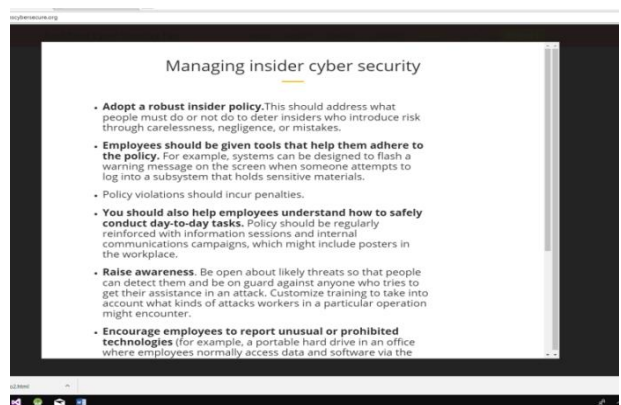

Figure 7: The Features Page with links to Cyber Security Tips

Figure 8: Cyber Security Tips Page

Details of one of the links in the features page that highlight how to plan against data insecurity by managing insider cyber security policies in an organization are represented in figure 8.

## 5. Conclusions

This study has initiated a pioneering effort on real time cyber security awareness by providing some data and information security tips online via the portal www.ganamoscybersecure.org. A review of the level of awareness of individuals and organizations on cyber security as well as their attitudes towards securing their data in the cloud was made to serve as a pointer to what to populate in the design. Also the strategies adopted by cyber criminals to succeed have been reviewed to provide information on how to guide against falling prey to hackers. It has been observed that majority of individuals and organizations are unaware of the risks their data and information are exposed to in communication networks and this also informs why most of them devote little attention to adopting data/information security strategies, thus making them prone to cyber attacks.

Technology alone cannot secure data and information; it has to be complimented by education and enlightenment of users of information technology resources and the practitioners alike. This is why most organizations are seeking the services of information security professionals who can create and implement a coprehensive information security program as well as provide support to make their employees information security conscious. It is therefore recommended that the outcome of this study and many more related efforts should be propagated to create adequate awareness on the menace of cyber insecurity and offer prevention services.